\documentclass[aps,prl,twocolumn,showpacs,superscriptaddress,groupedaddress]{revtex4}  % for review and submission
\usepackage{graphicx}  % needed for figures
\usepackage{dcolumn}   % needed for some tables
\usepackage{bm}        % for math
\usepackage{amssymb}   % for math

\begin{document}

% the following line is for submission, including submission to the arXiv!!
\hspace{5.2in} 
% I commented out:
%\mbox{Fermilab-Pub-04/xxx-E}

\title{Femtosecond compression dynamics and timing jitter suppression in a terahertz-driven electron bunch compressor}

\author{E. C. Snively}
\email{esnively@slac.stanford.edu}
\affiliation{SLAC National Accelerator Laboratory, 2575 Sand Hill Road, Menlo Park, California, 94025, USA}

\author{M. A. K. Othman}
\email{mothman@slac.stanford.edu}
\affiliation{SLAC National Accelerator Laboratory, 2575 Sand Hill Road, Menlo Park, California, 94025, USA}

\author{M. Kozina}
\affiliation{SLAC National Accelerator Laboratory, 2575 Sand Hill Road, Menlo Park, California, 94025, USA}

\author{B. K. Ofori-Okai}
\affiliation{SLAC National Accelerator Laboratory, 2575 Sand Hill Road, Menlo Park, California, 94025, USA}
%\affiliation{PULSE Institute, SLAC National Accelerator Laboratory, Menlo Park, CA 94025, USA}

\author{S. P. Weathersby}
\affiliation{SLAC National Accelerator Laboratory, 2575 Sand Hill Road, Menlo Park, California, 94025, USA}

\author{S. Park}
\affiliation{SLAC National Accelerator Laboratory, 2575 Sand Hill Road, Menlo Park, California, 94025, USA}
\affiliation{Stanford Institute for Materials and Energy Sciences, SLAC National Accelerator Laboratory, Menlo Park, CA 94025, USA}

\author{X. Shen}
\affiliation{SLAC National Accelerator Laboratory, 2575 Sand Hill Road, Menlo Park, California, 94025, USA}

\author{X. J. Wang}
\affiliation{SLAC National Accelerator Laboratory, 2575 Sand Hill Road, Menlo Park, California, 94025, USA}

\author{M. C. Hoffmann}
\email{hoffmann@slac.stanford.edu}
\affiliation{SLAC National Accelerator Laboratory, 2575 Sand Hill Road, Menlo Park, California, 94025, USA}

\author{R. K. Li}
\email{lirk@tsinghua.edu.cn}
\affiliation{SLAC National Accelerator Laboratory, 2575 Sand Hill Road, Menlo Park, California, 94025, USA}
\altaffiliation[Now at: ]{Tsinghua University, 30 Shuangqing Rd, Beijing Shi, China, 100091}

\author{E. A. Nanni}
\email{nanni@slac.stanford.edu}
\affiliation{SLAC National Accelerator Laboratory, 2575 Sand Hill Road, Menlo Park, California, 94025, USA}

\begin{abstract}
We present the first demonstration of THz-driven bunch compression and timing stabilization of a few-fC relativistic electron beam with kinetic energy of 2.5 MeV using quasi-single-cycle strong field THz radiation in a shorted parallel-plate structure. Compression by nearly a factor of 3 produced a 39 fs rms bunch length and a reduction in timing jitter by more than a factor of 2, to 31 fs rms, offering a significant improvement to beam performance for applications like ultrafast electron diffraction. This THz-driven technique provides a critical step towards unprecedented timing resolution in ultrafast sciences and other accelerator applications using femtosecond-scale electron beams.
\end{abstract}

% I commented out:
%\pacs{}
\maketitle

Advances in electron-beam based ultrafast science continue to reach unprecedented sensitivity, with techniques like pump-probe ultrafast electron diffraction (UED) achieving sub-angstrom spatial resolution and temporal resolutions down to 100 fs using electron bunch lengths of only tens of femtoseconds \cite{UEDSLACWeathersby,UEDMannebach,UEDYang}. The intense demand for ever shorter high-brightness electron beams has ignited a campaign to achieve fs-scale bunch lengths and timing stability through beam-wave interactions at THz frequencies \cite{SRRBaum,THzIFELCurry,STEAMZhang,THzwakefieldZhao,tiltedTHzBaum}, hand-in-hand with the advent of THz-based accelerator technology \cite{compressionTHzwaveguideWong,THznanotipWimmer,THzlinacNanni,THzgunHuang}. THz-driven compression techniques in the relativistic regime, where space-charge effects are suppressed, could potentially enable production of high-brightness ultra-short bunches with sufficient charge to capture single shot images in UED, allowing characterization of irreversible processes at the time scale of atomic motion \cite{UEDreview,UEDfutureDOE}. Bunch compression techniques using conventional radio frequency (RF) structures have succeeded in producing ultralow emittance few-fs beams \cite{deflectorMaxson} and represent a significant step towards achieving the high-brightness beams needed for single-shot UED. However, the interaction suffers from RF phase jitter, on the order of tens to hundreds of femtoseconds depending on the source stability \cite{jitterstudy,fsjitter,UEDjitter}, which can exacerbate the time-of-arrival jitter between compressed electron beam and reference pump laser. 

THz-driven beam manipulation has been recognized as a promising candidate in the pursuit of few-fs beams with sub-fs timing resolution, because the inherent timing synchronization of all-optical control enables both bunch compression and reduction of beam timing jitter. Already, THz-driven beam manipulation has demonstrated dramatic compression in the sub-relativistic regime ($<$100 keV electrons), reaching bunch lengths of tens of femtoseconds while reducing the timing jitter to a few femtoseconds \cite{SRRBaum,STEAMZhang}. Beyond the synchronization benefit that comes with a laser-driven interaction, structures operating in the THz regime offer a host of advantages. For applications like bunch compression or transverse deflection, the higher frequency provides a more efficient time-dependent momentum kick compared to conventional RF manipulation. Using sub-wavelength structures for localized field enhancement \cite{Fabianska}, THz-driven streaking diagnostics have already demonstrated femtosecond, down to sub-femtosecond metrology of relativistic beams \cite{streakingZhao,streakingLi}. Additionally, the sub-mm length scale of the THz regime enables the use of small-footprint structures supporting strong synchronism for efficient beam manipulation while occupying only mm-scale space on the beamline.

\begin{figure}[!htb]
\centering
\includegraphics*[width=.48\textwidth]{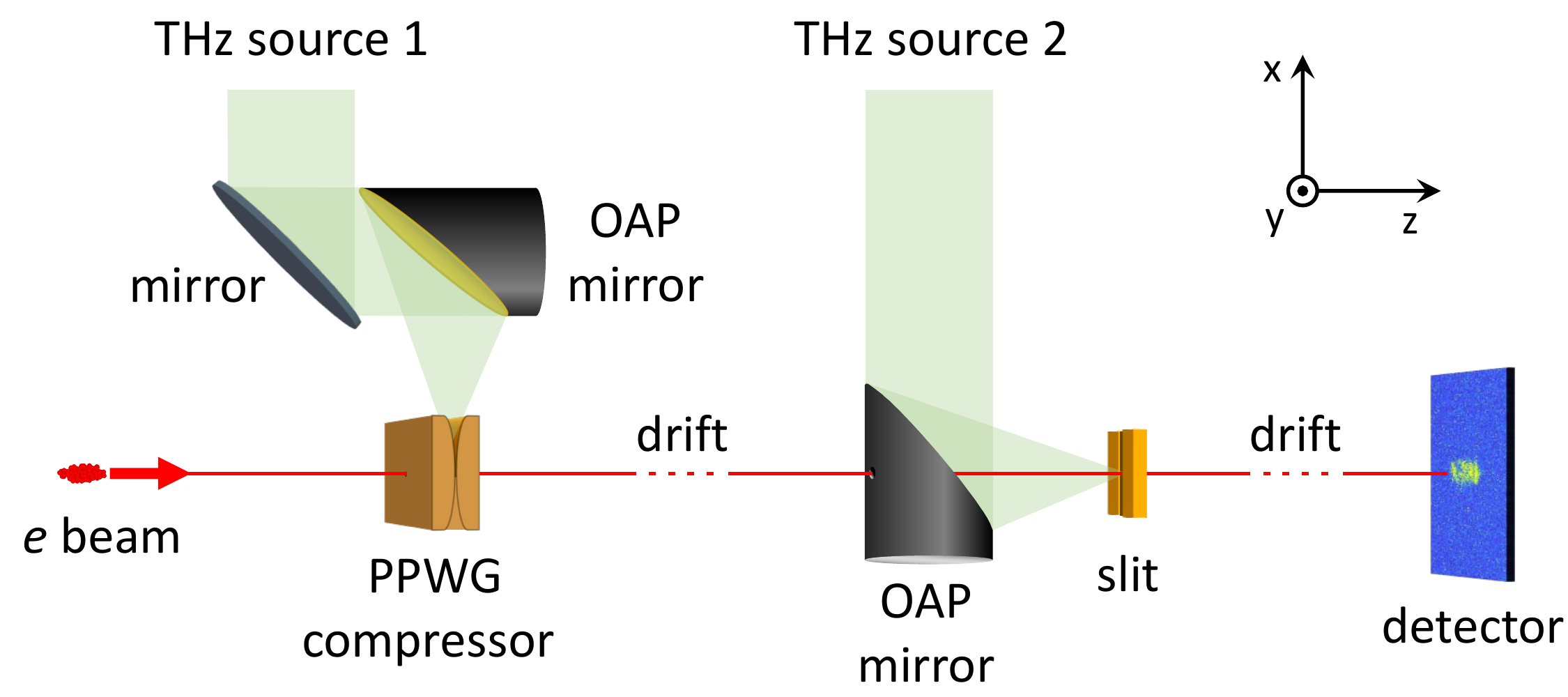}
\caption{Schematic of the in-vacuum components for the THz-driven compression and streaking setup. As the electron beam passes through the parallel plate waveguide (PPWG) compressor, the orthogonally propagating THz pulse imparts an energy chirp, resulting in velocity bunching during the subsequent drift. The beam then passes through an OAP mirror into collinear propagation with a second THz pulse as it is focused into a metallic slit. Here, the THz pulse imparts a transverse momentum kick to the beam which streaks the longitudinal profile onto the y-axis on a downstream imaging detector. THz source 1 is polarized along the z-axis; THz source 2 is polarized along the y-axis.}
\label{setup_diagram}
\end{figure}

We present the results of a THz-driven compression experiment performed at the SLAC MeV-UED facility at SLAC National Accelerator Laboratory using a shorted parallel-plate waveguide (PPWG) design \cite{compressorOthman}. Both the manipulation and characterization of the relativistic electron beam were accomplished through interaction with quasi-single-cycle THz pulses generated via optical rectification of 800 nm laser pulses \cite{THzASTAOfori}. Using the shorted PPWG to couple the relativistic electron beam and an orthogonally propagating $<\,3\,\mu$J THz pulse, we produced a beam energy chirp resulting in compression by a factor of 2.7, with a minimum bunch length of 39$\pm$7 fs rms. An equally important consequence of this interaction is the simultaneous improvement to the beam's shot-to-shot time-of-arrival (rms) stability by a factor of $>2.5$, in this case reducing the timing jitter to 31 fs rms. This THz-driven compression experiment builds on the success of the THz-based streaking diagnostic developed at SLAC MeV-UED, which previously demonstrated sub-fs rms timing accuracy of few MeV beams \cite{streakingLi}. The components of the THz-driven bunch compression and timing stabilization setup are integrated with the THz-driven beam diagnostic, as shown in Fig. \ref{setup_diagram}. The enhanced capabilities of these combined technologies are uniquely positioned to be directly applied to UED experiments.
% A statement should be added here to emphasize the nolveity of our work compared to [8]

The Ti:Sapphire laser system at SLAC MeV-UED provided a 13$\pm$1 mJ, near-IR (800 nm) pulse at 180 Hz which was split to drive the UV source, for electron beam generation in the S-band photocathode gun, and two separate THz sources, in which the pulse-front-tilt method was used for optical rectification in LiNbO$_3$ crystals \cite{PFTHebling,mJTHzFulop}. A 70-30 beam splitter directed the primary IR pulse to THz source 1, dedicated to driving bunch compression, with the secondary pulse continuing on to THz source 2 for the streaking setup. Two translation stages controlled the timing of each THz source via adjustment of the IR path length. To produce the horizontally polarized THz field required for interaction in the compressor, the components of the corresponding pulse-front-tilted optical rectification setup were mounted to a vertical breadboard. The resulting quasi-single-cycle THz pulse was transported into a vacuum chamber on the beamline using a pair of off-axis parabolic mirrors (OAP) to first collimate and then focus the pulse into the horn of the compressor structure, as shown in Fig. \ref{setup_diagram}. Downstream, the second THz source produced a quasi-single-cycle pulse polarized along the y-axis, which was similarly collimated and transported into a second vacuum chamber using a pair of OAPs. In this case, the final OAP was mounted on the beamline axis such that a small 2.5 mm hole in the mirror allowed the electron beam to pass through into co-propagation with the THz pulse as it was focused into a metallic PPWG ``slit" structure for streaking the beam \cite{streakingLi}. 

The 2.856 GHz photocathode gun on the SLAC MeV-UED beamline supplied a relativistic electron beam with 2.5 MeV kinetic energy and sub-10 fC total charge. After initial collimation with a focusing solenoid, the beam was aligned through a pinhole and then focused with a second solenoid to a $\sim40 \,\mu$m rms spot size in the compressing structure. The bunch length and timing jitter of the uncompressed beam were 105$\pm$19 fs rms and 76 fs rms, respectively, measured by the THz-driven streaking diagnostic 1 m downstream of the compressor.     

\begin{figure}[!b]
\centering
\includegraphics*[width=.48\textwidth]{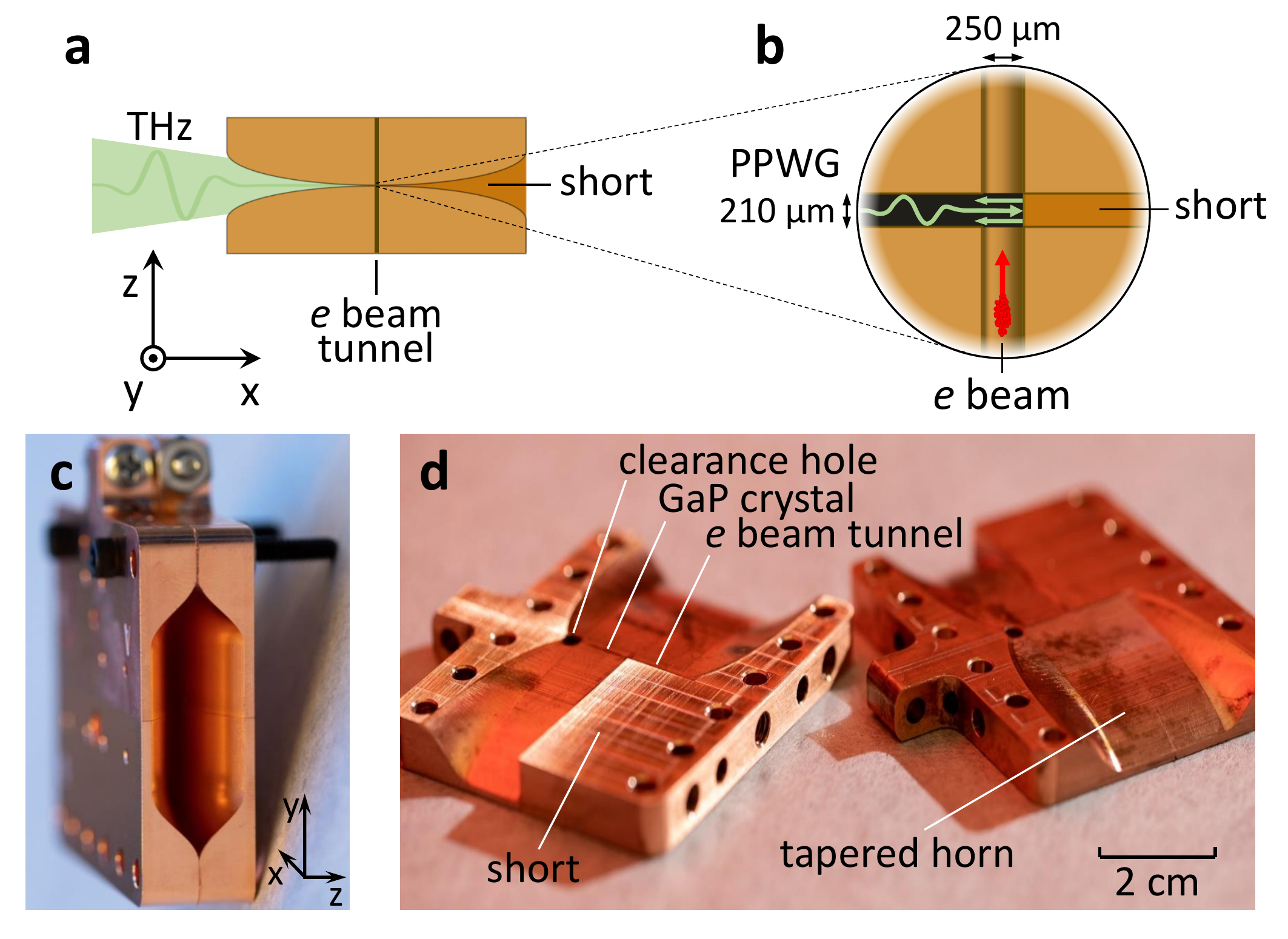}
\caption{(a) Diagram of the compressor structure cross-section, showing the exponentially tapered horn with incoming THz pulse polarized along the z-axis. (b) A close-up view of the interaction region, showing the 250 $\mu$m diameter beam tunnel and the PPWG with 210 $\mu$m gap. A short reflects the incoming THz pulse as the beam passes through. (c) A photograph of the assembled structure showing the adiabatic horn opening. (d) A photograph of the disassembled structure before the final etch.  The structure is translated along the y-axis to switch between compressor ``on" with beam passing through the beam tunnel and compressor ``off" with beam passing through the 2.54 mm diameter clearance hole. A GaP crystal, mounted between the beam tunnel and clearance hole, enabled in-situ electro-optic sampling of the THz field.}
\label{compressor_diagram}
\end{figure}

The THz-driven compression technique presented here utilized a PPWG structure that benefits from two key design enhancements. Where the THz pulse entered the structure, an exponentially-tapered adiabatic horn was matched to the free-space THz beam profile, significantly improving the coupling efficiency and allowing dispersion-free focusing of the THz pulse into the PPWG at the center of the structure. Using the coordinate system of the compressor diagram in Fig. \ref{compressor_diagram}, the adiabatic horn focused the z-component of the quasi-single-cycle THz pulse as it propagated along the x-axis, overcoming the free space diffraction limit and resulting in significant field enhancement between the parallel plates aligned in the x-y plane. Within the 210 $\mu$m gap of the PPWG structure, the addition of a copper short, located at the edge of the beam tunnel (125 $\mu$m from beam axis), produced a superposition of forward and reflected THz field in the vicinity of the passing electron beam, increasing the parallel electric field driving the energy chirp, while reducing the magnetic field which imparted an undesirable transverse momentum kick. 

\begin{figure}[!htb]
\centering
\includegraphics*[width=.48\textwidth]{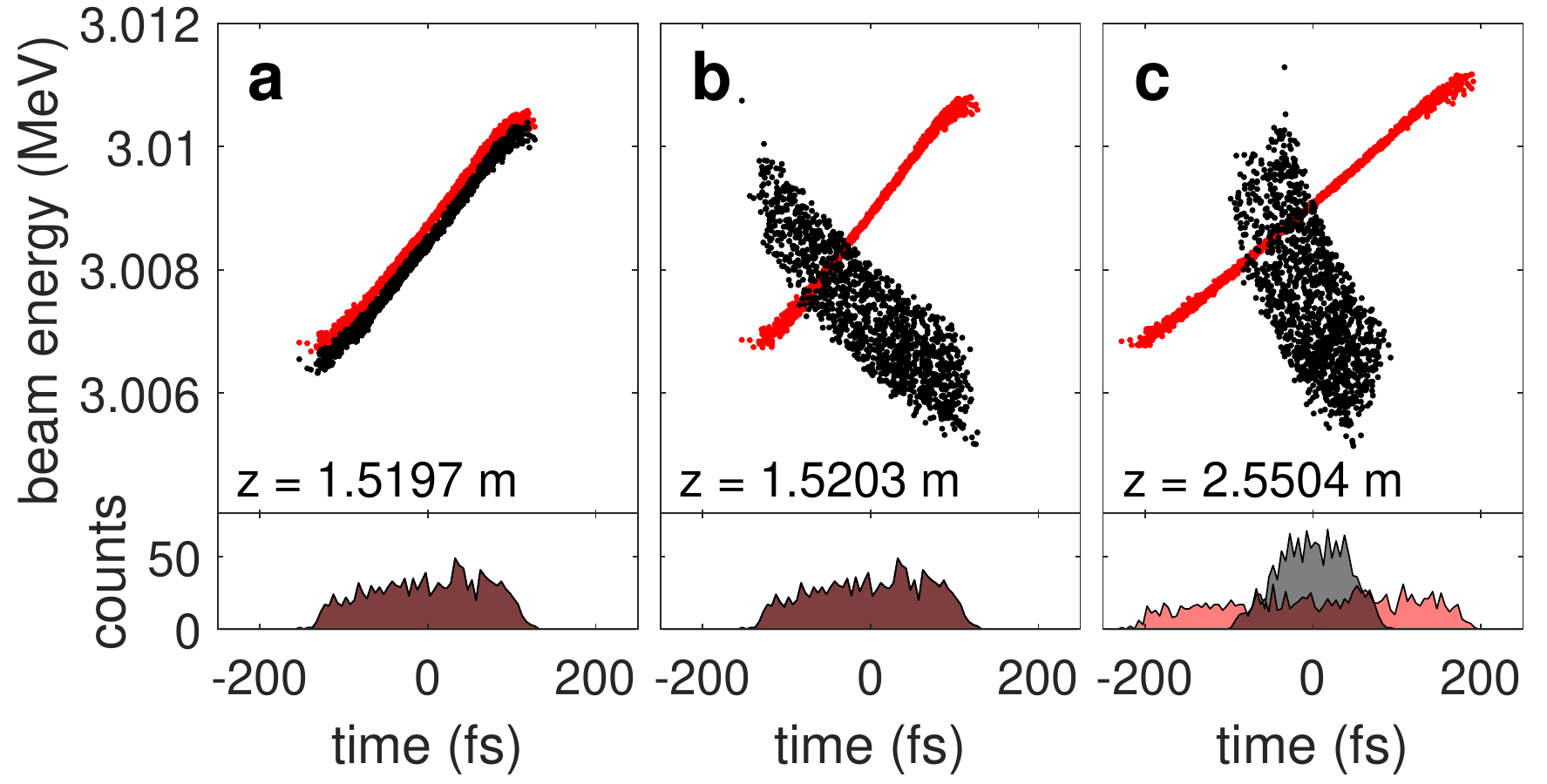}
\caption{Simulated longitudinal phase space of the beam at different points along the beamline, with compressor ``off" shown in red and compressor ``on" in black. (a) Just before entering the compressor, the two cases show the same distribution. (b) As it nears the exit of the compressor, the beam that interacted with the THz pulse (black) has accumulated an energy chirp of opposite sign. During the drift, higher energy electrons move forward relative to the center of the bunch, while lower energy electrons move back. (c) As a result, downstream at the streaking slit, the uncompressed beam (red) has increased in length, while the compressed beam (black) has become shorter through velocity bunching. The projected bunch length is shown at the bottom of each plot.}
\label{GPT_LPS}
\end{figure}

As the electron beam traversed the PPWG gap, the integrated Lorentz force experienced by particles at different positions along the bunch length was determined by the temporal overlap with the THz field. Ideally, the beam was injected at a phase with maximum THz electric field gradient, typically near a zero-crossing in the waveform, which produced the strongest near-linear longitudinal energy chirp, resulting in efficient velocity bunching of the beam. The temporal profile of the THz pulse determined the phase acceptance window over which the beam received an energy chirp with the correct sign for bunch compression. For beams arriving within this window, the THz electric field gradient increased the energy of particles at the tail of the beam relative to particles at the head of the beam, as shown in Fig. \ref{GPT_LPS}b. It follows directly that within this window, the energy of a ``late" bunch was increased relative to an ``early" bunch, such that the compressor interaction also reduced the shot-to-shot beam timing jitter after a subsequent drift. This effect takes advantage of the inherent timing synchronization between the THz pulse and electron beam generated using the same initial laser pulse.

In addition to producing the energy chirp necessary for velocity bunching, the Lorentz force within the PPWG interaction region produced a time-dependent transverse deflection of the beam. The shorted PPWG configuration was chosen to reduce this transverse deflection and defocusing caused by the magnetic field, with the superposition of the forward and reflected THz wave producing constructive interference of the transverse electric field while the magnetic field was partially canceled \cite{compressorOthman}. At the local maxima in deflection, the shot-to-shot beam pointing jitter increased to 30 $\mu$rad rms from the 10 $\mu$rad rms jitter without THz interaction. The interaction phase for maximum compression, near a local maximum in deflection, produced a shot-to-shot pointing jitter of up to 50 $\mu$rad rms. Steering magnets placed after the compression chamber were used to compensate for the compressor-induced deflection, ensuring consistent alignment through the streaking diagnostic.

We observed both compression of the electron bunch length and a reduction in the shot-to-shot timing jitter using the THz-driven streaking setup positioned 1 m downstream from the compression interaction. Here the longitudinal profile of the electron beam was projected onto the vertical axis through the time-dependent transverse momentum kick imparted to the beam by a co-propagating THz pulse in a metallic slit \cite{streakingLi}. The direction and magnitude of the deflection was determined by the temporal profile of the THz pulse. A calibration of femtoseconds per pixel was directly obtained by scanning the streaking THz pulse arrival time and mapping the corresponding deflection of the beam centroid on the phosphor screen placed 2 m downstream of the streaking interaction.

To extract the beam characteristics from a single shot image of the charge profile after streaking, we compare the measured distribution to the distribution produced by assuming a Gaussian longitudinal profile for a test beam which is then mapped into a ``projected distribution" using the known streaking calibration curve. The length and arrival time of the Gaussian test beam are allowed to vary in order to find the best fit of the simulated distribution to the actual projected beam image. This method effectively deals with the irregular charge distributions produced when the initial longitudinal beam profile fills the region mapped out by the THz-driven transverse deflection, causing the charge to ``pile-up" at the ends of the projected distribution.  

\begin{figure*}[!htb]
\centering
\includegraphics[width=1\textwidth]{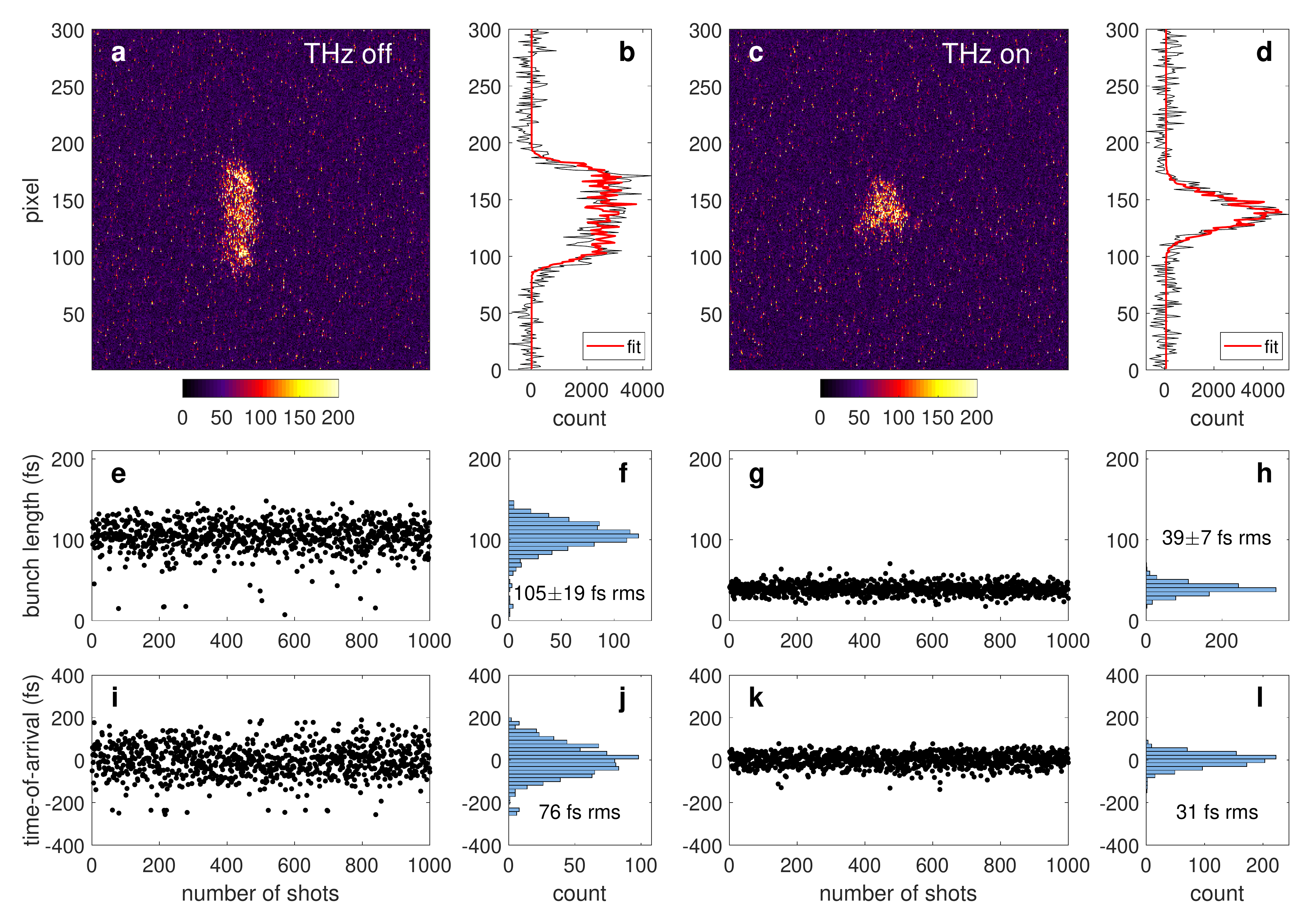}
\caption{Comparison of the uncompressed beam to the maximally compressed beam. Single shot beam images in (a) and (c) show examples with THz-driven compression ``off" and ``on", respectively, with the corresponding fit to their projected distribution shown in (b) and (d). (e)-(f) show the bunch length of the uncompressed beam over 1000 shots, with an average bunch length of 105$\pm$19 fs rms. (g)-(h) show the bunch length of the compressed beam over 1000 shots, indicating an average bunch length of 39$\pm$7 fs rms. (i)-(j) show the time-of-arrival of the uncompressed beam over 1000 shots, with a timing jitter of 76 fs rms. (k)-(l) show the time-of-arrival of the compressed beam over 1000 shots, giving a reduced timing jitter of 31 fs rms.}
\label{on_off_example}
\end{figure*}

\begin{figure*}[!htb]
\centering
\includegraphics[width=1\textwidth]{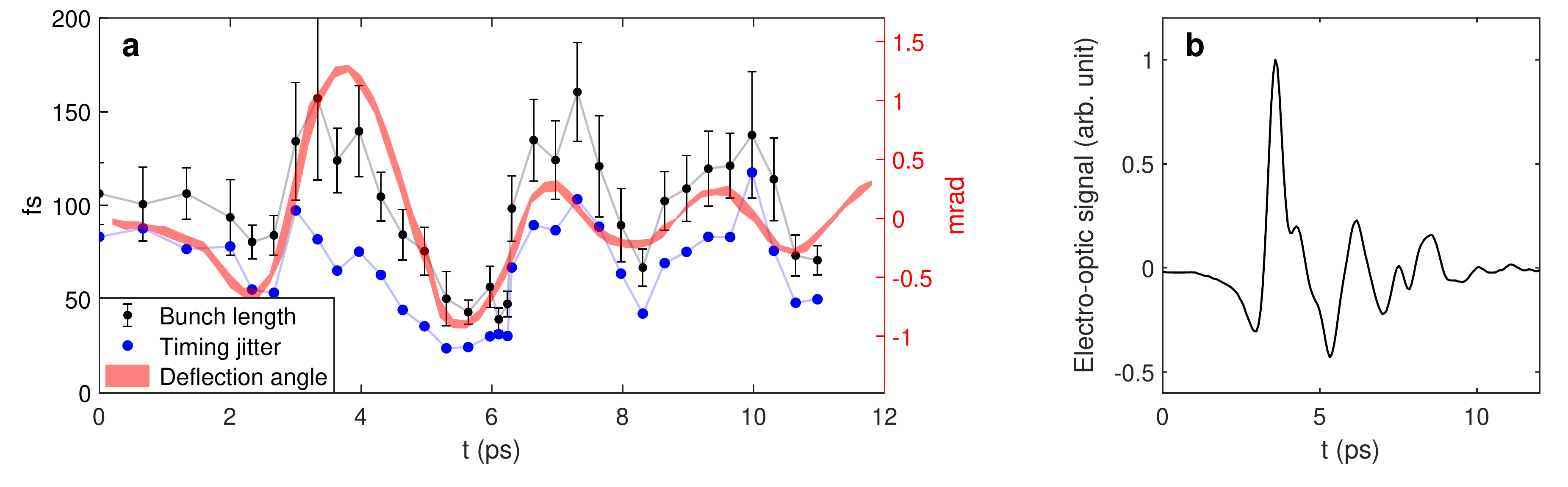}
\caption{(a) Scan of the bunch length and timing jitter of the electron beam, along with the deflection angle in the x-z plane (scale shown on right axis), as a function of the THz pulse arrival time within the structure. The quasi-single-cycle profile of the driving THz field is evident in the time-dependent oscillations of the beam characteristics, each showing a parallel change in sign and magnitude. (b) Temporal profile of the THz pulse measured within the PPWG structure via electro-optic sampling with the GaP crystal. This measurement characterizes the THz field in the un-shorted region of the PPWG structure. The ringing apparent after $\sim$6 ps is an aftifact from the EO sampling crystal, not the true waveform.}
\label{comp_scan}
\end{figure*}

In Figure \ref{on_off_example}a-d, we show examples of single shot beam images alongside their corresponding projected distribution, with the background subtracted, and the fit produced through the test beam method. These examples highlight two cases, in which the compression interaction is first turned off and then turned on to produce a maximally compressed bunch. In Fig. \ref{on_off_example}e-l, we plot the bunch length and time-of-arrival for 1000 shots of the uncompressed beam, indicating a mean bunch length of 105$\pm$19 fs rms and timing jitter of 76 fs rms, and for 1000 shots of the compressed beam, indicating a mean bunch length of 39$\pm$7 fs rms and timing jitter of 31 fs rms. The reduction in timing jitter by a factor of 2.5 is slightly smaller than the measured compression factor of 2.7. However, similar measurements of 1000 shots taken at neighboring interaction phases near the minimum bunch length show timing jitters as low as 24 fs rms, as shown in Fig. \ref{comp_scan}a, indicating jitter reduction by up to a factor of 3.2.

In Fig. \ref{comp_scan}a, we plot the results of a scan of THz arrival time within the compressing structure, including the corresponding bunch length, timing jitter and transverse deflection. These parameters vary in time according to the quasi-single-cycle profile of the driving THz field. The deflection data shown in Fig. \ref{comp_scan}a was collected in a separate scan, independent of the streaking setup, with finer temporal resolution. The width of the line reflects the standard deviation in beam centroid position along the x-axis in the lab frame. Because the compressor-induced transverse deflection of the beam necessitated realignment between each streaking measurement, the full set of bunch length and timing jitter measurements was acquired over a couple hours. During this time, a slow drift in the relative time-of-arrival resulted in a cumulative scaling uncertainty on the order of 1 ps per hour. 

The profile of the THz pulse measured via EO sampling in the un-shorted section of the PPWG is shown in Fig. \ref{comp_scan}b. These EO measurements were used to generate a simulated profiled of the THz field in the shorted PPWG section with Ansoft HFSS software \cite{compressorOthman}. Simulations of the electron beam dynamics were performed with the GPT simulation package \cite{GPT}, beginning in the photocathode gun and including the compressor field maps imported from HFSS. Given our measured beam energy and uncompressed bunch length of 105 fs rms, GPT simulation shows that the peak field sampled by the beam within the compressor must be around 100 MV/m in order to reproduce the compressed bunch length of 39 fs rms as measured at the streaking slit. This value is consistent with our expected field enhancement, given the THz pulse energy and structure design \cite{compressorOthman}. These simulated values correspond to an incoming beam energy chirp of 1.7 keV per 100 fs, and exiting chirp of -1.3 keV per 100 fs, and indicate that maximum compression of the 2.5 MeV bunch occurred approximately 70 cm downstream of the streaking slit, where the beam reached a minimum bunch length of $\sim$32 fs rms. The actual bunch compression achieved in our experiment was primarily limited by the available THz energy, with the measurement of that compression further limited by the location of our streaking diagnostic. Eventually, the compression achieved through this shorted PPWG technique would also be limited by the increase to the slice energy spread of the beam. 

The transverse deflection evident in our measurements and in GPT simulations of the compressor interaction is predominantly along the x-axis. The measured deflection along the y-axis mirrored the x-deflection but with smaller magnitude, reaching only 250 $\mu$rad where the x-deflection peaked to 1.27 mrad. The magnitude of the measured y-deflection is consistent with simulations, but the GPT model predicts x-deflection that is a factor of two larger than observed in measurements. Differences in the position of the beam within the tunnel cross-section are a likely source of discrepancy between our measurements and simulation, causing the beam to sample a different region of the shorted field profile compared to the on-axis trajectory assumed in simulation. This inhomogeneity caused by the short in the PPWG structure is key to the reduction in the transverse deflection, and required careful optimization of the distance between the short and beam axis for the phase of the forward and reflected magnetic field to add destructively along the beam path \cite{compressorOthman}. 

A more straight-forward design for mitigating the transverse momentum kick is a dual feed structure that utilizes two counterpropagating THz pulses. This technique has been successfully demonstrated for THz-driven acceleration, compression and streaking of a sub-relativistic beam \cite{STEAMZhang}. While this method offers advantages in the flexibility and fidelity of the THz field superposition, it introduces the additional challenge of requiring either a second THz source, or THz splitting optics with additional beam transport. In the development of next-generation single-feed PPWG structures, a split waveguide design, in which the beam tunnel passes through two PPWG interaction regions, could provide added flexibility for optimization of the energy chirp while canceling the induced deflection. This alternate design could also improve the uniformity of the field in the vicinity of the beam, reducing the slice energy spread growth predicted for the current design. 

In summary, we have shown a compression interaction, driven by a laser-generated THz pulse in a parallel plate waveguide, that reduces the bunch length while acting to correct the timing jitter of the relativistic beam, stabilizing its time-of-arrival with respect to a reference laser pulse. The simultaneous compression to a 39 fs rms bunch length and 31 fs rms timing jitter provides a significant benefit to performance for electron-beam based ultrafast science, given that the overall temporal resolution of a measurement is dependent on the bunch length and timing jitter added in quadrature. The transverse deflection of the beam induced by this compression interaction is easily corrected using steering magnets and introduces only a minimal increase to the beam pointing jitter. The magnitude of compression achieved through this technique could be directly improved by increasing the input THz pulse energy, with state-of-the-art THz sources already providing greater pulses energies by more than an order of magnitude \cite{FulopmJTHz,THzDASTVicario}. This demonstration of THz-driven compression and timing stabilization is a critical step towards achieving the ultra-short electron beams that could enable UED measurements at the attosecond scale. More broadly, electron beams conditioned through this technique are advantageous for a range of accelerator applications, like pump-probe ultrafast electron scattering and external injection in laser-driven accelerators that require few-fs bunch lengths and timing stability \cite{plasmaAccelMalka,plasmaAccelEsarey,DLAEngland}.  

\begin{acknowledgments}

The authors acknowledge M. Cardoso, A. Sy, and A Haase for structure fabrication. This research has been supported by the U.S. Department of Energy (DOE) under Contract No. DE-AC02-76SF00515. The SLAC MeV-UED program is supported in part by DOE Basic Energy Sciences (BES) Scientific User Facilities Division and SLAC UED/UEM program development: DE-AC02-05CH11231. M. K. and M. C. H. are supported by the DOE Office of Science, BES, award no. 2015-SLAC100238-Funding. 

E. C. S., M. A. K. O., and M. K. contributed equally to this work.

\end{acknowledgments}

\end{document}